\documentstyle[12pt,aaspp4,flushrt]{article}

\def\ang{~{\rm \AA}}
\def\cm2{~\rm cm^{-2}}
\def\hno{{\rm H}_0}
\def\kms{~{\rm km\ s}^{-1}}
\def\lal{{\rm Ly}\alpha}
\def\lsim{\buildrel < \over \sim}
\def\msun{~\rm M_{\odot}}
\def\qno{q_0}
\def\ten#1{\times 10^{#1}}

\begin{document}

\title{A Lower Limit to the Universal Density of Metals at $z \sim 3$}
\author{Antoinette Songaila\altaffilmark{1}} 
\affil{Institute for Astronomy, University of Hawaii, 2680 Woodlawn Drive,
  Honolulu, HI 96822\\}

\altaffiltext{1}{The author was a visiting astronomer at the W. M. Keck
  Observatory, jointly operated by the California Institute of Technology and
  the University of California.}

\vskip 1in

\centerline{Accepted for publication in {\it Astrophysical Journal Letters\/}}

\begin{abstract}

Column density distribution functions of \ion{C}{4} with $12.05 \le \log (N)
\le 14.35$\ and \ion{Si}{4} with $11.70 \le \log (N) \le 13.93$\ have been
obtained using 81 \ion{C}{4} absorbers and 35 \ion{Si}{4} absorbers redward
of the $\lal$\ forest in the lines of sight to seven quasars with $2.518 \le
z_{em} \le 3.78$.  These distribution functions have been directly integrated
to yield ion densities at $z = 3$\ to 3.5 of $\Omega_{\rm C~IV} = (2.0 \pm 0.5)
\ten{-8}$\ and $\Omega_{\rm Si~IV} = (7.0 \pm 2.6) \ten{-9}$\ with $\hno =
65\kms\ {\rm Mpc}^{-1}$\ and $\qno = 0.02$\ ($1~\sigma$\ errors).  A larger
sample of 11 quasar lines of sight was used to measure \ion{C}{2}/\ion{C}{4},
\ion{Si}{3}/\ion{Si}{4}, and \ion{N}{5}/\ion{C}{4} ratios, which suggest that
\ion{C}{4} and \ion{Si}{4} are the dominant ionization stages and that
corrections to $\Omega_{\rm Carbon}$\ and $\Omega_{\rm Silicon}$\ are no more
than a factor of two.  Normalizing the $\alpha$-process elements to silicon
and the Fe-coproduction elements to carbon gives a density of heavy elements
in these forest clouds of $\Omega_{\rm metals} = (3.3 \pm 0.8)\ten{-7}$\
($\hno = 65,\ \qno = 0.02)$.  The implications for the amount of star
formation and for the ionization of the IGM prior to $z = 3$\ are discussed.

\end{abstract}


\section{Introduction} \label{intro}

In any CDM or ${\rm CDM} + \Lambda$\ cosmology, the first generation of stars
is expected to form at redshifts 10--30 in small ($\sim 10^8\msun$) bound
objects and is assumed to imprint the intergalactic medium by both enriching
and ionizing it.  This enrichment may now have been detected in Keck spectra
of the Lyman forest (\markcite{llc95}Cowie et al.\ 1995;
\markcite{tyt95}Tytler et al.\ 1995; \markcite{sc96}Songaila \& Cowie 1996
[SC96]), and various attempts have been made to use the first crude
measurements of the forest metallicity (\markcite{sc96}SC96;
\markcite{hsr96}Haehnelt, Steinmetz, \& Rauch 1996; \markcite{rhs97}Rauch,
Haehnelt \& Steinmetz 1997) to place upper limits on the amount of early star
formation in small galaxies (\markcite{sc96}SC96;
\markcite{mr97}Miralda-Escud\'e \& Rees 1997; \markcite{go97}Gnedin \&
Ostriker 1997; \markcite{hl97}Haiman \& Loeb 1997) and to infer the rate of
the high redshift type--II supernovae that may be our best hope of detecting
objects in these early stages of the galaxy formation era
(\markcite{mr97}Miralda-Escud\'e \& Rees 1997).

The critical quantity for normalizing the amount of early star formation is
the density of metals at early epochs.  This has generally been obtained (e.g.
\markcite{mr97}Miralda-Escud\'e \& Rees 1997; \markcite{hl97}Haiman \& Loeb
1997) by combining the rough estimates of metallicity in forest clouds with
the total baryon density ($\Omega_b$) inferred from big bang nucleosynthesis
(SBBN).  Apart from the uncertainty in the value of $\Omega_b$\ from SBBN,
this methodology assumes a uniform enrichment of metals in the IGM and also
entails considerable uncertainty in the determination of individual cloud
metallicities caused by poorly known ionization corrections and abundance
patterns in the heavy elements.

It is possible to avoid many of these problems by directly integrating the
observed ion column densities in the forest clouds to obtain $\Omega_{\rm
ion}$\ and, within plausible limits, $\Omega_{\rm metals}$.  Carbon and
silicon are both well suited to this approach: as discussed below, their
dominant ionization stages in the IGM are \ion{C}{4} and \ion{Si}{4}, strong
doublets that are easy to find and measure with 10 m class telescopes
(\S\ref{obs}); they both have strong accessible lines in other ionization
stages with which to assess ionization corrections (\S\ref{ion}); and the
ratio of $\alpha$-process silicon to Fe-coproduction carbon is useful for
determining the abundance pattern at early epochs and to give some idea of the
nature of the initial mass function (\S\ref{discuss}).  This {\it Letter\/}
reports the results of such a direct integration.

\section{Observations} \label{obs}

The data used comprises high resolution observations of eleven quasars with
$2.518 \le z_{em} \le 3.78$\ made with the HIRES spectrograph on the Keck I
telescope, for a variety of programs, between 1994 November and 1997 April.
All spectra have a resolution $R = 36,000$\ and have variable wavelength
coverage in the range $3500\ang - 7000\ang$.  Full details of the extraction
and processing of these spectra are given in a companion paper
(\markcite{song97}Songaila 1997).  Seven of the spectra have complete
wavelength coverage between the quasar's $\lal$\ and \ion{C}{4} emission lines,
whereas the remainder have only partial coverage redward of $\lal$\ emission or
have remaining inter-order gaps from HIRES's incomplete coverage above
$5000\ang$.

The sample used to construct the \ion{C}{4} column density distribution
consists of all \ion{C}{4} doublets detected redward of $\lal$\ emission in
those spectra with complete coverage.  The doublets were found by inspection
of the spectra, and confirmed by consistency of the column density and
velocity structure in the two members of the doublet.  Systems within
$5000\kms$\ of the quasar's emission redshift were excluded from the sample
to avoid including proximate systems which have ion ratios dominated by
photoionization from the quasars themselves.  Also excluded were 10
specifically targeted partial Lyman limit systems (PLLS) in 9 quasars that
were chosen to be observed in a program to look for suitable systems in which
to measure the primordial ratio of deuterium to hydrogen.  The final sample
consists of 81 \ion{C}{4} doublets, with $2.02 \le z \le 3.54$\ and $12.05
\le \log (N_{\rm C~IV}) \le 14.35$, free from proximity or observational
selection bias.

The \ion{Si}{4} column density distribution was similarly determined from all
\ion{Si}{4} doublets detected in the same subsample of quasar spectra as used
for the \ion{C}{4} distribution.  Only systems blueward of the quasars'
rest-frame \ion{Si}{4} but redward of the Lyman forest were included, again
excluding proximate systems and targeted PLLS.  This resulted in 35 systems
in seven lines of sight between $z = 2.16$\ and $z = 3.54$, with $11.70 \le
\log (N_{\rm Si~IV}) \le 13.93$.

In all cases, ion column densities were determined by fitting up to ten Voigt
profiles to each redshift system, defined in this context to be all
absorption near a given fiducial redshift with gaps in velocity space of no
more than $50\kms$.  The column density at a given redshift is then the total
column density of all such components.  In all but a few systems, individual
lines were unsaturated, so fitted column densities are insensitive to
$b$-value.

With these samples, the column density distribution function $f(N)$\ was
determined for each ion.  $f(N)$\ is defined as the number of absorbing
systems per unit redshift path per unit column density, where at a given
redshift $z$\ the redshift path, $X(z)$, is defined as $X(z) \equiv
\case{1}{2}\, [(1+z)^2 - 1]$\ for $\qno = 0$\ or by $X(z) \equiv \case{2}{3}\,
[(1+z)^{3/2} - 1]$\ for $\qno = 0.5$ (\markcite{tyt82}Tytler 1982).  The 
\ion{C}{4} and \ion{Si}{4} distribution functions are shown in 
Figure~1, over the redshift ranges, $2.02 \le z \le 3$\
and $3 \le z \le 3.54$\ for \ion{C}{4}, and $2.16 \le z \le 3$\ and $3 \le z \le
3.54$\ for \ion{Si}{4}.  The data were calculated with $\qno = 0$\ and are
plotted with $1~\sigma$\ error bars calculated from the Poisson errors based
on the number of systems in each bin.

Because of the somewhat heterogeneous nature of the parent observations, the
signal-to-noise of the final sample is quite variable, in some cases because
the low quasar emission redshift entailed observing below $4000\ang$\ where
the CCD efficiency dropoff is quite severe, in others because of variable
coverage at the red end of the spectrum to fill in inter-order gaps.
Exposure times were also variable for the usual observational reasons.  The
amount of incompleteness at low column density was assessed by recalculating
the column density distributions using only systems drawn from the highest
signal-to-noise spectrum, toward Q1422+231, which has an exposure time of 580
minutes and reaches a $1~\sigma$\ limiting column density of $1.3
\ten{11}\cm2$\ for \ion{C}{4} and $5\ten{10}\cm2$\ for \ion{Si}{4} for
$b=6\kms$.  Comparing this with the distributions obtained from the full
sample, it is found that the turnover at low column density ($N_{\rm C~IV} <
2.8\ten{12}\cm2$) in the full sample is almost entirely a result of
incompleteness.  To take this into account, power laws were fitted only above
$5.6\ten{12}\cm2$\ for \ion{C}{4} and $10^{12}\cm2$\ for \ion{Si}{4} , with
best-fit indices of $-1.5$\ for both low and high redshift \ion{C}{4}, and
$-1.8$\ for low redshift \ion{Si}{4} and $-2.0$\ for high redshift
\ion{Si}{4}.  These values are quite similar to those measured in \ion{H}{1}
in the forest at these redshifts (\markcite{petit93}Petitjean et al.\ 1993;
\markcite{hu95}Hu et al.\ 1995; \markcite{kim97}Kim et al.\ 1997) and in
higher column density \ion{C}{4} samples (\markcite{petit94}Petitjean \&
Bergeron 1994).

$\Omega_{\rm C~IV}$\ and $\Omega_{\rm Si~IV}$\ were calculated from the column
density distributions of Figure~1 according to
\begin{equation}
\Omega_{\rm ion} = {{\hno} \over {\rho_{\rm crit}}} \  
{{\sum N_{\rm ion}} \over {\Delta X}} \  m_{\rm ion} \label{eq:omega_ion}
\end{equation}
where $\rho_{\rm crit} = 1.89 \ten{-29} h^2~{\rm g\ cm}^{-3}$\ is the
cosmological closure density, $m_{\rm ion}$\ is the ion's mass, and $\hno =
100\,h\kms\ {\rm Mpc}^{-1}$ (e.g., \markcite{lanz91}Lanzetta et al.\ 1991).
Values of $\Omega_{\rm C~IV}$\ and $\Omega_{\rm Si~IV}$\ were calulated for 
$z > 3$\ in individual lines of sight to the five quasars in the sample with
$z_{em} > 3$\ and complete spectral coverage between the quasar's
$\lal$\ and \ion{C}{4} emission lines.  The results are tabulated in
Table~\ref{tbl:1} for $\hno = 65\kms~{\rm Mpc}^{-1}$\ and $\qno =
0.02$\ ($\Omega_{\rm ion}$\ scales as $h^{-1}\, [1+2\qno z]^{1/2}$) and give
some idea of the uncertainty in calculating $\Omega_{\rm ion}$.  Formal mean
values, weighted by $\Delta X$, are $\Omega_{\rm C~IV} = (1.2 \pm 0.3)
\ten{-8}\,h^{-1}\, [1+2\qno z]^{1/2}$\ and $\Omega_{\rm Si~IV} = (4.3 \pm 1.6) \ten{-9}\,h^{-1}\, [1+2\qno z]^{1/2}$ , where the
errors are $1~\sigma$.  The mean redshift of the sample is $\langle
z\rangle=3.18$.  A similar procedure applied to the $z<3$ systems gives
$\Omega_{\rm C~IV} = 9.3\ten{-9}\,h^{-1}\, [1+2\qno z]^{1/2}$ and $\Omega_{\rm
Si~IV} = 5.0\ten{-9}\,h^{-1}\, [1+2\qno z]^{1/2}$ at $\langle z\rangle=2.57$.

Because of the rarity of high column density systems ($N_{\rm C~IV} \gg
3\ten{14}\cm2$) considerably longer path lengths are required to determine the
number density of such systems, and the present data measure $\Omega_{\rm
ion}$\ only for absorption with column density less than this value.  (This
corresponds roughly to $N_{\rm H~I} < 10^{17}\cm2$, from \markcite{sc96}SC96.)
The metal densities of the stronger systems are addressed in a companion paper
(\markcite{song97}Songaila 1997).  The contribution of systems to $\Omega_{\rm
C~IV}$\ and $\Omega_{\rm Si~IV}$\ converges at the low column density end, and
systems weaker than those observed will not contribute significantly to the
density unless there is a very rapid upturn below the observed range.

\section{Ionization Balance} \label{ion}

A less restricted sample was used to assess the ionization balance correction
to be applied to the \ion{C}{4} and \ion{Si}{4} sample.  This was drawn from
the full set of eleven quasar spectra, excluding proximate systems.  In cases
where there was no apparent absorption at the \ion{C}{4} redshift in the
other ionic species, an upper limit to the absorption in other ions was found
by formally fitting the \ion{C}{4} component model to the local continuum.
This approach results in conservative upper limits, especially to lower
ionization species, since \ion{C}{4} absorption is likely to be more
widespread in velocity space than lower ionization absorption.

As is shown in Figure~2, \ion{C}{2} is a trace ion relative to
\ion{C}{4} with nearly all systems having \ion{C}{2} $\ll$\ 0.1~\ion{C}{4}.  An
additional direct search in the spectra for \ion{C}{2} systems with no strong
\ion{C}{4} yielded no additional systems.  For photoionization models, this in
turn yields a high photoionization parameter ($\log \Gamma \gg -2$) which, for
a wide range of photoionizing spectra, implies that \ion{Si}{3} $\lsim$
\ion{Si}{4} and \ion{C}{3} $\lsim$ \ion{C}{4}  (e.g., \markcite{st90}Steidel
1990; \markcite{gs97}Giroux \& Shull 1997).  For \ion{Si}{3} this can be
directly verified from observations (middle panel of Figure~2) which
show that, even with $\lal$\ forest contamination, the upper limits on the 
\ion{Si}{3} column density are generally compatible with \ion{Si}{4} measurements.
Ionization by a starburst galaxy spectrum with no high energy photons could
result in a much higher ratio of \ion{C}{3} to \ion{C}{4} while maintaining the
low \ion{C}{2}/\ion{C}{4} and \ion{Si}{3}/\ion{Si}{4} ratios, but it would also
produce \ion{Si}{4} much in excess of \ion{C}{4}, which is not observed:  at $z
> 3$\ \ion{Si}{4}/\ion{C}{4} has an average value of about 0.2
(\markcite{sc96}SC96; \markcite{song97}Songaila 1997).

Finally, neither \ion{N}{5} nor \ion{O}{6} is strong in the
\ion{C}{4}-selected systems.  The right panel of Figure~2 shows
that \ion{N}{5} $<$ 0.05 \ion{C}{4} in the small number of systems measured.
The high observed ratio of \ion{Si}{4} to \ion{C}{4} combined with the low
observed values of \ion{C}{2}/\ion{C}{4} and \ion{N}{5}/\ion{C}{4} suggests
that forest clouds at $z > 3$\ are ionized by a broken power low spectrum
with relatively few high energy photons and that it is unlikely that there is
much material above the \ion{C}{4} and \ion{Si}{4} levels
(\markcite{sc96}SC96; \markcite{gs97}Giroux \& Shull 1997).  The best guess,
therefore, would give $\Omega_{\rm carbon} \lsim 2\,\Omega_{\rm C~IV}$\ and
$\Omega_{\rm silicon} \lsim 2\,\Omega_{\rm Si~IV}$\ in these clouds.

\section{Discussion} \label{discuss}

Converting $\Omega_{\rm carbon}$\ and $\Omega_{\rm silicon}$\ to the
$\Omega_{\rm metals}$\ of all metals requires an assumption of an abundance
pattern in the $z > 3$\ forest.  As in the old metal-poor halo stars, silicon
is overabundant with respect to carbon relative to solar.  However, assuming
that the carbon and silicon abundances trace the universal abundances of iron
coproduction and $\alpha$-process elements, respectively (e.g.,
\markcite{timmes95}Timmes, Lauroesch, \& Truran 1995), then for the
Fe-coproduction elements (C, N, Fe), $\Omega = 1.75 \times \Omega_{\rm C}$,
and for the $\alpha$-process elements (O, Ne, Si, Mg, S), $\Omega = 18.5
\times \Omega_{\rm Si}$, and setting $\Omega_{\rm C} = 2 \Omega_{\rm C~IV}$
and $\Omega_{\rm Si} = 2 \Omega_{\rm Si~IV}$ gives $\Omega_{\rm metals} = (2.1
\pm 0.6)\ten{-7}\,h^{-1}\, [1+2\qno z]^{1/2}$ (relative abundances from
\markcite{ag89}Anders \& Grevesse 1989).  Assuming a value of $\Omega_b$\ from
SBBN of $0.005 \le \Omega_b \, h^2 \le 0.016$\ (\markcite{swc97}Songaila,
Wampler, \& Cowie 1997) gives $\Omega_{\rm metals}/\Omega_b = (1.3 \pm 0.4
\rightarrow 4.2 \pm 1.2)\ten{-5}\, h\, [1+2\qno z]^{1/2}$.  With $\hno =
65\kms\ {\rm Mpc}^{-1}$, $\qno = 0.02$, and a solar metallicity of 0.019, this
implies a minimum universal metallicity relative to solar, at $z
\sim 3$, in the range [$-3.3$] to [$-2.8$].  
(The value of $D/H = 2.3\ten{-5} \pm 3\ten{-6}\ ({\rm statistical}) \pm
3\ten{-6}\ ({\rm systematic})$\ given in \markcite{tfb96}Tytler, Fan \& Burles
(1996) has now been revised upwards to $3.3\ten{-5}$\ (\markcite{tyt97}Tytler
1997) which would imply $\Omega_b\,h^2 = 0.016$\ and a metallicity of [$-3.3$],
while the value of $2\ten{-4}$\ obtained by \markcite{webb97}Webb et al.\
(1997) would give a metallicity of [$-2.9$] [$\hno = 65,\ \qno = 0.02$].)
This is a minimum range of
metallicity since there may be additional metals in higher column density
clouds or galaxies as well as in ionization states that were not sampled by
these observations.  Furthermore, adoption of $\qno = 0.5$ would roughly
double the metallicity.  However, much of $\Omega_b$\ inferred from SBBN is
believed to reside in the forest clouds at this time (\markcite{rauch}Rauch
et al.\ 1997; \markcite{kim97}Kim et al.\ 1997), suggesting that this
calculated metal density should be a good estimate of the total metals.
Interestingly, since a metallicity of $10^{-5}$\ gives rise to
one ionizing photon per baryon (e.g., \markcite{mr97}Miralda-Escud\'e \& Rees
1997), the metallicity measured here is just sufficient to preionize the
IGM.  However, this metallicity range is substantially lower than the value
of $2.4\ten{-4}$\ assumed by \markcite{mr97}Miralda-Escud\'e \& Rees (1997)
and \markcite{hl97}Haiman \& Loeb (1997) and implies that their predicted
rate of very high-$z$\ type-II supernovae could be lowered by as much as an
order of magnitude, to a value of around 1 SN per 10 arcmin$^2$\ per year.

\acknowledgments
I am grateful to the many people at the Keck telescopes who made these
observations possible, and to Len Cowie and Esther Hu for obtaining some of
the observations on which this work is based.  The research was supported by
the National Science Foundation under grant AST 96-17216.

\newpage

\begin{deluxetable}{ccccccc}
\tablewidth{450pt}
\tablecaption{$z > 3\ \ \Omega_{\rm ion}$\ by Quasar \label{tbl:1}}
\tablehead{
\colhead{Quasar} & \colhead{$z_{em}$} & \colhead{$\Delta X$} &
\colhead{$\sum N({\rm C IV})/\Delta X$} & \colhead{$\sum N({\rm Si IV})/\Delta
X$} & \colhead{$\Omega_{\rm C IV}$} &  \colhead{$\Omega_{\rm Si IV}$} \\
[0.5ex] &&& ($\times 10^{14}\cm2$) & ($\times 10^{13}\cm2$) & ($\times
10^{-8}$) &  ($\times 10^{-9}$)
}
\startdata
0014+813   & 3.384 & 1.29 & 0.62 & 0.42 & 1.2   & 1.8  \nl
0302$-$003 & 3.286 & 0.88 & 0.51 & 0.17 & 0.9   & 0.74 \nl
0956+122   & 3.301 & 0.94 & 2.4  & 3.7  & 4.5   & 16.0 \nl
1159+123   & 3.493 & 1.76 & 0.62 & 0.63 & 1.2   & 2.7  \nl
1422+231   & 3.620 & 2.32 & 1.3  & 2.7  & 2.4   & 12.0 \nl
\tablevspace{15pt}
Mean&&& $(1.1 \pm 0.3)$ &  $(1.6 \pm
0.6)$ & $(2.0 \pm 0.5)$ & $(7.0 \pm 2.6)$ \nl
\enddata
\tablecomments{For all five quasars, \ion{Si}{4} and \ion{C}{4} lie above the
$\lal$\ forest at $z > 3$\ so $\Delta X$\ is the same for \ion{C}{4} and
\ion{Si}{4}.  $\Omega$\ is computed for $\hno = 65\kms~{\rm Mpc}^{-1}$\ and 
$\qno = 0.02$. Mean values are weighted by $\Delta X$\ and quoted errors are 
$1~\sigma$.}
\end{deluxetable}
\clearpage

\newpage

\begin{figure}
\figurenum{1(a)}
\plotone{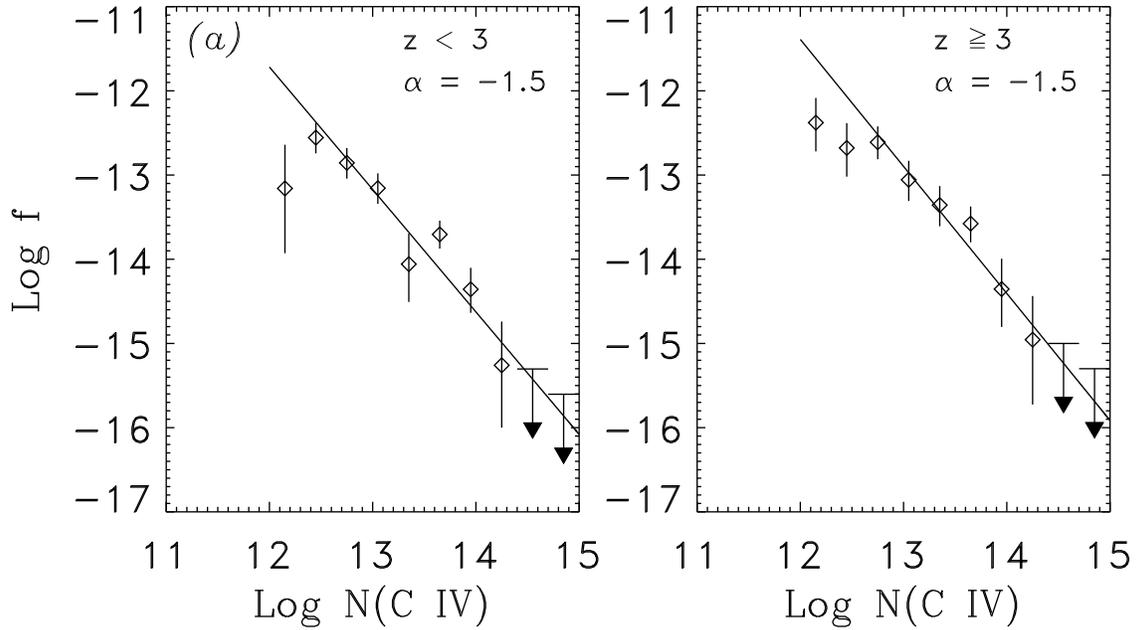}
\caption{C~IV column density 
distribution functions for $z < 3$\ and $z \ge 3$\ for a total of 81 absorption
line systems in seven lines of sight toward distant quasars.  The bin size is
$10^{0.3}\, N~{\rm cm}^{-2}$.  Error bars are $\pm 1~\sigma$, based on the number
of systems in each bin.  The solid lines show the best fit power law with index
$\alpha = -1.5$, fitted to data above $6\times 10^{12}~{\rm cm}^{-2}$ where the
data is substantially complete.}

\end{figure}

\begin{figure}
\figurenum{1(b)}
\plotone{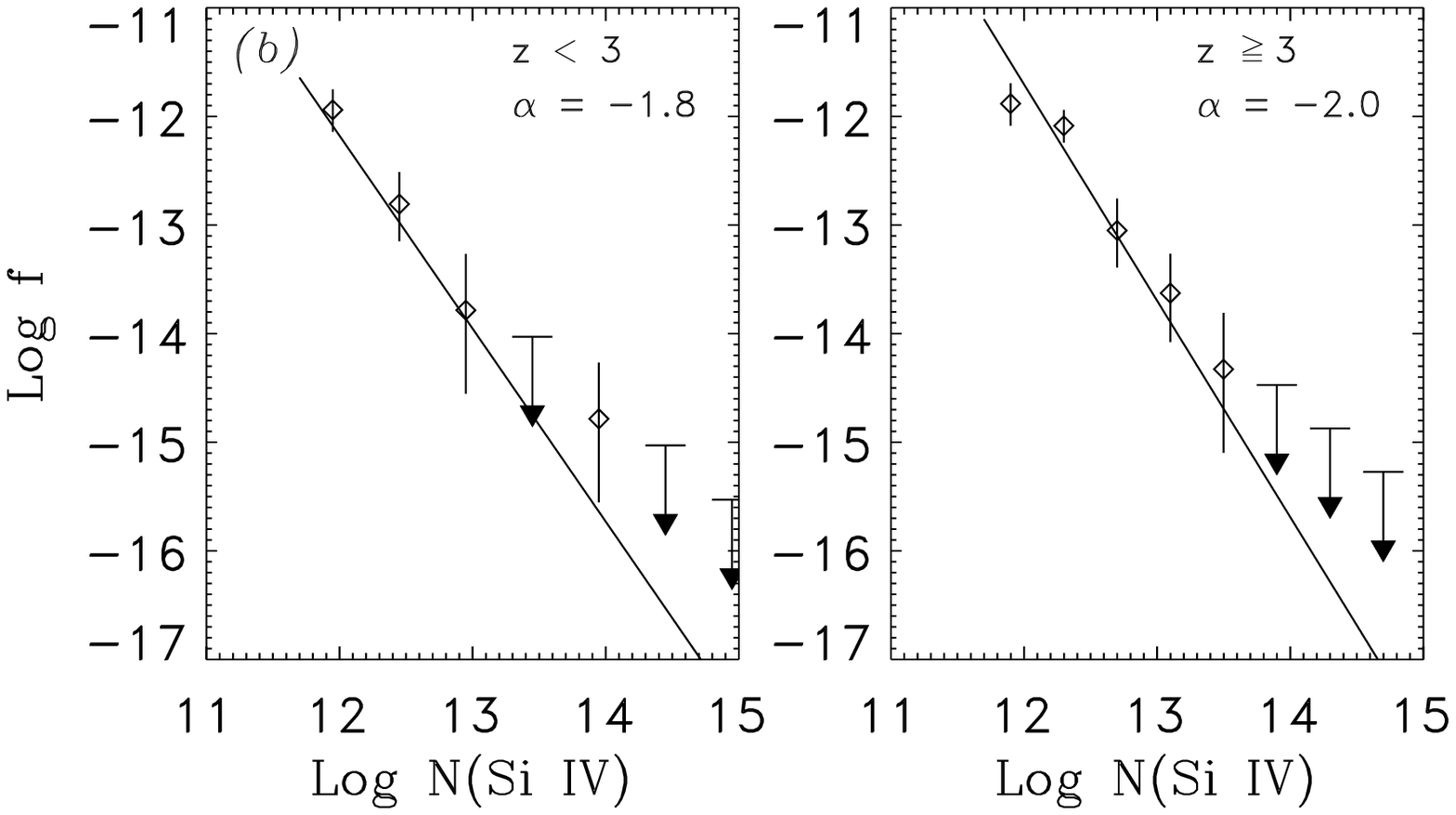}
\caption{As in (a) for Si~IV, with a total
of 35 systems and a bin size of $10^{0.5}\, N~{\rm cm}^{-2}$\ for $z < 3$\ and
$10^{0.4}\, N~{\rm cm}^{-2}$\ for $z \ge 3$. The solid lines are power laws of
index $\alpha = -1.8$\ for $z < 3$\ and $\alpha = -2$\ for $z \ge 3$,
fitted to data between $10^{11}~{\rm cm}^{-2}$\ and $10^{14}~{\rm cm}^{-2}$.}
\end{figure}

\begin{figure}
\figurenum{2}
\plotone{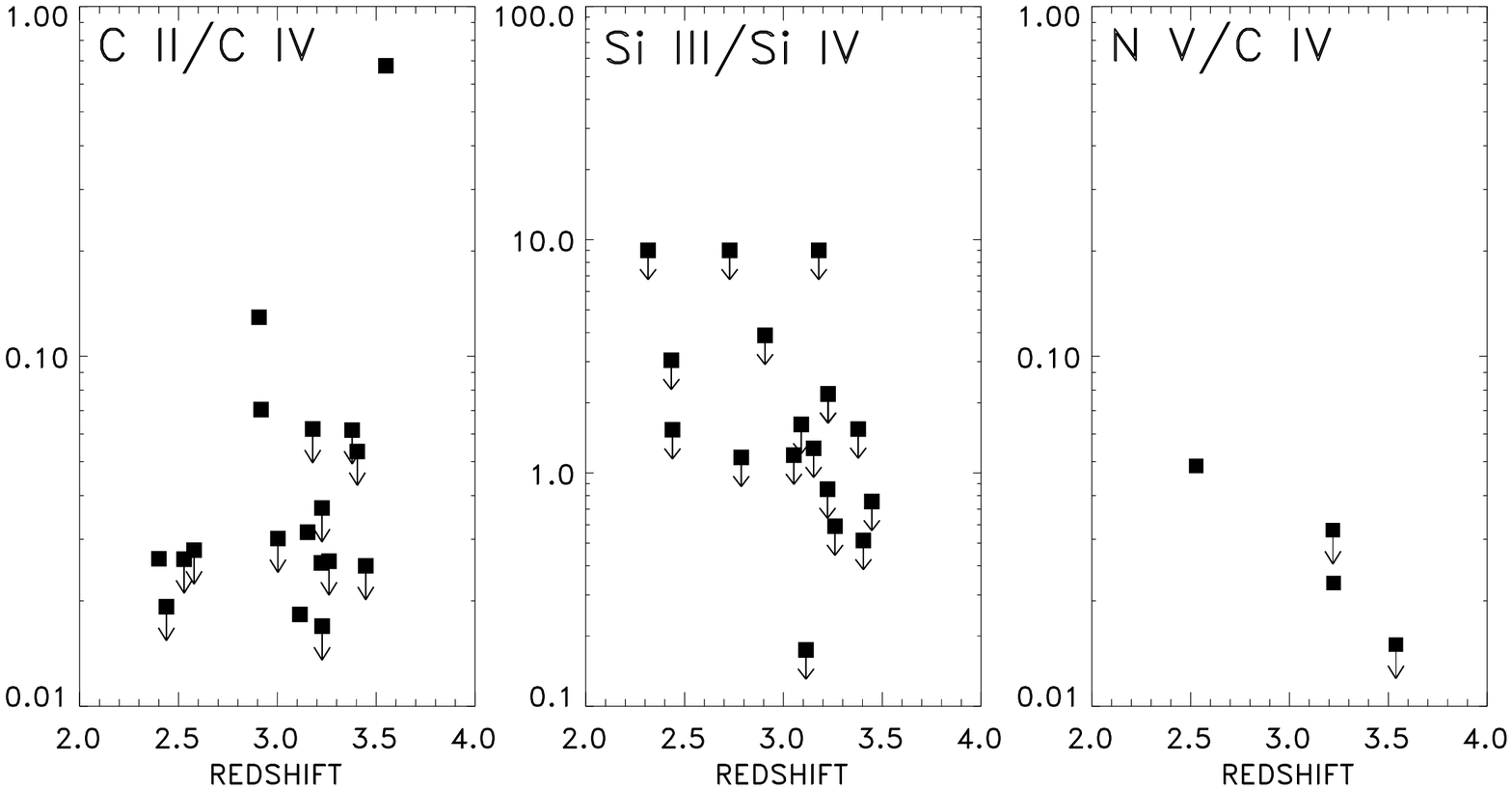}
\caption{Ion ratios for absorption line 
systems toward the full quasar sample.  The C~II and N~V samples were
C~IV-selected ($N_{\rm C~IV} > 1.5\times 10^{13}~{\rm cm}^{-2}$\ for C~II;
$N_{\rm C~IV} > 4\times 10^{12}~{\rm cm}^{-2}$\ for N~V) and the Si~III sample
was Si~IV-selected ($N_{\rm Si~IV} > 2\times 10^{12}~{\rm cm}^{-2}$).  For C~II
and N~V, the sample was restricted to lines lying longward of the
Ly$\alpha$\ forest.  This is not possible for ${\rm Si~III}\ \lambda
1206~{\rm\AA}$, where all measured values were treated as upper limits because
of forest contamination.}
\end{figure}

\end{document}